# High Performance *In Vivo* Near-IR (>1 μm) Imaging and Photothermal Cancer Therapy with Carbon Nanotubes

Joshua T. Robinson[1], Kevin Welsher[1], Scott M. Tabakman[1], Sarah P. Sherlock[1], Hailiang Wang[1], Richard Luong[2], and Hongjie Dai[1] (✉)

[1] Department of Chemistry, Stanford University, Stanford, CA 94305, USA
[2] Department of Comparative Medicine, Stanford University School of Medicine, Stanford, CA 94305, USA

## ABSTRACT

Short single-walled carbon nanotubes (SWNTs) functionalized by PEGylated phospholipids are biologically non-toxic and long-circulating nanomaterials with intrinsic near infrared photoluminescence (NIR PL), characteristic Raman spectra, and strong optical absorbance in the near infrared (NIR). This work demonstrates the first dual application of intravenously injected SWNTs as photoluminescent agents for *in vivo* tumor imaging in the 1.0–1.4 μm emission region and as NIR absorbers and heaters at 808 nm for photothermal tumor elimination at the lowest injected dose (70 μg of SWNT/mouse, equivalent to 3.6 mg/kg) and laser irradiation power (0.6 W/cm$^2$) reported to date. *Ex vivo* resonance Raman imaging revealed the SWNT distribution within tumors at a high spatial resolution. Complete tumor elimination was achieved for large numbers of photothermally treated mice without any toxic side effects after more than six months post-treatment. Further, side-by-side experiments were carried out to compare the performance of SWNTs and gold nanorods (AuNRs) at an injected dose of 700 μg of AuNR/mouse (equivalent to 35 mg/kg) in NIR photothermal ablation of tumors *in vivo*. Highly effective tumor elimination with SWNTs was achieved at 10 times lower injected doses and lower irradiation powers than for AuNRs. These results suggest there are significant benefits of utilizing the intrinsic properties of biocompatible SWNTs for combined cancer imaging and therapy.

## 1. Introduction

Single-walled carbon nanotubes (SWNTs) have shown promise for biological applications [1] due to their ability to load targeting ligands [2] and chemotherapy agents [3, 4]. By virtue of their high optical absorbance in the biological transparency window of ~0.8–1.4 μm, SWNTs may also act as photothermal therapy agents [5, 6]. The intrinsic optical properties of SWNTs, such as strong resonance Raman scattering and near infrared photoluminescence (NIR PL) in the 1.1–1.4 μm spectral region, make them useful biological imaging agents [7–15]. Imaging of nanoparticle uptake into tumors, through the enhanced permeability and retention (EPR) effect [16], is key to nanomaterial-based cancer therapeutics. Nanoparticles, such as nanotubes and gold nanoparticles, have been used as NIR contrast agents [17, 18]. Previously, *ex vivo* and *in vivo* spectroscopy,

Address correspondence to hdai@stanford.edu

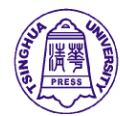

such as Raman scattering, have both been used to evaluate SWNT uptake [2, 13]. Other *in vivo* methods have also been utilized, such as positron emission tomography (PET) for radio-labeled nanotubes [2]. Photoacoustic imaging [19] was used to monitor SWNT tumor uptake *in vivo*, but whole animal imaging has not been accomplished.

The inherent NIR PL of semiconducting SWNTs has proven useful as a biological imaging modality, due to the excitation being in the biological transparency window [18], the large Stokes shift between excitation and emission, and ultralow autofluorescence background and deep tissue penetration in the emission range (1.1–1.4 μm) [14, 15, 20]. SWNTs have been used as PL imaging tags inside macrophage cells [8] and in *Drosophila melanogaster* (fruit flies) *in vivo* [10] based on their NIR PL. The ability to use SWNTs as fluorescent agents for targeted *in vitro* cell imaging with high specificity has been demonstrated [12]. The intrinsic photoluminescence of SWNTs has also been shown to be viable for whole body imaging and intravital tumor vessel imaging *in vivo* following intravenous injection [15].

A related area of research is photothermal therapy. When photosensitizers accumulate in close proximity in the body, a light source can be used to heat tissue to a level which results in photocoagulation and cell death [21]. Usually, this technique is used to reduce the size of, or eliminate, tumors. It has been difficult to find a method that achieves high tumor accumulation of the photosensitizers without direct injection into the tumor. Novel nanomaterials have shown promise for photothermal therapy due to their unique size and optical properties. Work has been done exploring nanomaterials as viable photothermal agents, such as gold nanoshells [21–24], gold nanorods (AuNRs) [25–28], gold nanopyramids [29], multi-walled carbon nanotubes (MWCNTs) [30], and SWNTs [31–33]. *In vivo* work has been done with intra-tumor injections of SWNTs, followed by excitation and heating by either radio frequencies or a NIR laser [6, 31, 33]. Intra-tumor injections require prior knowledge of the location of the tumor, whereas intravenous injection of a material with imaging and photothermal capabilities and high tumor uptake does not.

Here, we demonstrate high tumor uptake of SWNTs with passive (non-targeted) accumulation following intravenous (i. v.) injection, and elimination of tumor masses with unprecedented low NIR laser powers of only 0.6 W/cm$^2$. We also demonstrate the first NIR imaging of tumors using semiconducting SWNTs as photoluminescent agents *in vivo*. Tumor uptake of nanotubes is assessed *in vivo* using whole animal imaging of the NIR PL intrinsic to SWNTs. Following imaging, we utilize the high optical absorbance of the SWNTs to photothermally heat whole tumors, reaching a temperature (~52 °C) that leads to complete tumor destruction. In a comparative study, SWNTs exhibited higher tumor photoablation ability than AuNRs. While SWNTs at an injected dose of 3.6 mg/kg afforded tumor NIR ablation at a power of 0.6 W/cm$^2$, AuNRs, at an injected dose of ~35 mg/kg, required a power of 2 W/cm$^2$ for tumor elimination. Throughout the treatment process, no toxicity of the SWNTs was observed *in vivo*. These results demonstrate the high performance of SWNTs for *in vivo* tumor imaging and photothermal therapy without obvious toxic side effects.

## 2. Experimental

### 2.1 Functionalized carbon nanotubes and gold nanorods

In a typical experiment, HiPco SWNTs were functionalized and solubilized by a mixture of 50% 1,2-distearoyl-phosphatidylethanolamine-methyl-polyethyleneglycol (DSPE-mPEG) [1, 2] and 50% C$_{18}$-PMH-mPEG surfactants (Fig. 1(a)). C$_{18}$-PMH-mPEG, (poly(maleic anhydride-*alt*-1-octadecene)-poly(ethylene glycol) methyl ether), is a PEG-branched polymer capable of binding to SWNTs and affording ultralong blood circulation of nanotubes (half-life ~18.9 h) [34]. The resulting suspension was centrifuged at 22 000 *g* for 6 h to remove large aggregates and nanotube bundles [35]. The average length of the SWNTs was ~140 nm as determined by atomic force microscopy (AFM) (Fig. 1(c), Fig. S-2 in the Electronic Supplementary Material (ESM)). The NIR PL was characterized by photoluminescence excitation and emission (PLE) spectroscopy (Fig. 1(b)). The spots in the PLE map (Fig. 1(b)) correspond to excitation and emission of individual SWNTs with different chiralities [36].

Gold nanorods (AuNRs) were prepared via seed-mediated growth in cetyltrimethylammonium bromide (CTAB) as described previously [37]. The CTAB coating on the AuNRs was exchanged for a methyoxy-terminated, thiolated poly(ethylene glycol) (mPEG-SH) coating (Fig. 1(d)) and the solution concentrated to 3.5 mg/mL prior to injection, after removal of excess mPEG-SH by centrifugal filtration (molecular weight cut off (MWCO) 30 kDa). The size and shape of the AuNRs was confirmed by transmission electron microscopy (TEM, Fig. 1(e)). At the same mass concentration (0.35 mg/mL), SWNTs exhibited a three-fold higher optical absorbance at 808 nm, which is the wavelength commonly used for NIR heating using nanomaterials (Fig. 1(f)).

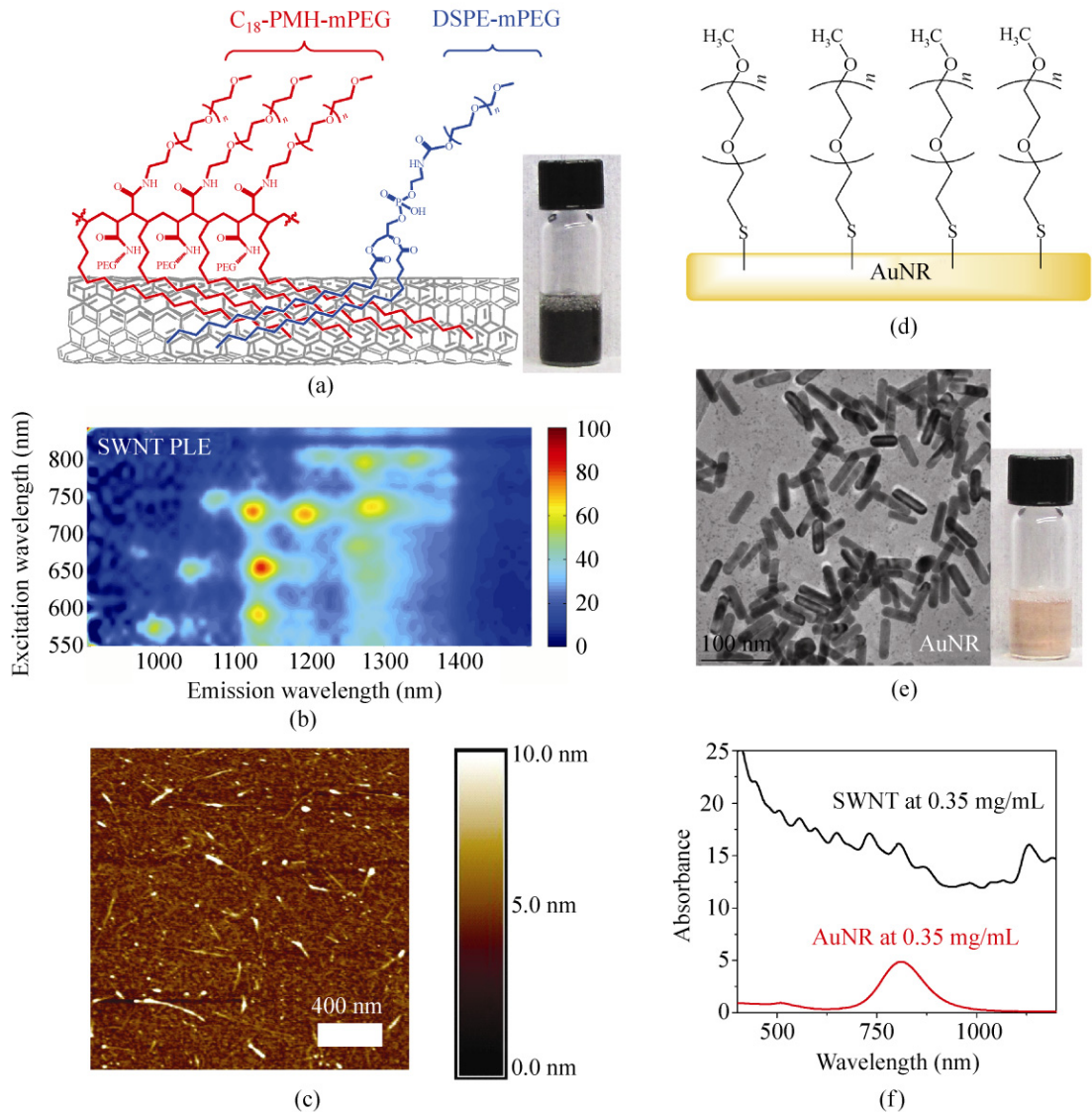

**Figure 1** Nanotubes and nanorods for biological applications. (a) Schematic illustration of the surface functionalization of an SWNT by $C_{18}$-PMH-mPEG and phospholipid DSPE-mPEG in a solubilized suspension. Inset: photo of a stable suspension of functionalized SWNTs in water. (b) PLE spectrum of a SWNT suspension. Individual peaks correspond to semiconducting SWNTs with different chirality. (c) An AFM topography image of SWNTs. The height variation along the nanotubes in the AFM likely reflects the large differences in size of the mixed surfactants used ($C_{18}$-PMH- mPEG and DSPE-PEG have molecular weights of ~1 MD and 5 kD respectively). (d) Schematic illustration of gold nanorods (AuNR) functionalized by mPEG-SH (5000 D). (e) A TEM image of AuNRs. Inset: photo of a stable suspension of functionalized AuNRs in water. (f) UV–vis–NIR absorption curves of SWNTs and AuNRs at an equal mass concentration of 0.35 mg/mL. A scatter component is superimposed for both nanoparticles

## 2.2 Tumor imaging using the intrinsic photoluminescence of SWNTs

Tumor-bearing mice were obtained by subcutaneous inoculation of ~2 million 4T1 murine breast tumor cells in BALB/c mice (Fig. S-2(a) in the ESM). For imaging, the typical SWNT injection dose was 200 μL of SWNTs at a concentration of 0.35 mg/mL through i. v. injection, corresponding to a dose of 3.6 mg/kg of mouse body weight. Different amphiphilic polymer coatings used to suspend SWNTs greatly affect their circulation time and biodistribution [38], including tumor uptake. By measuring the intrinsic Raman signals of SWNTs in blood and various organs (see the Methods section) [38], we found that the 50% DSPE-mPEG/50% $C_{18}$-PMH-mPEG coating gave SWNTs with a long blood circulation half-life of 6.9 h (Fig. 2(a)) and high tumor uptake of 8% ID/g (injected dose/gram of tissue), despite high uptake in the liver and spleen (Fig. 2(b)).

We were able to monitor SWNT fluorescence *in vivo* over the course of two days using a 2D InGaAs detector ($n > 10$) (Figs. 2(c) and 2(d)). Very low autofluorescence was observed from untreated mice in the collection window (1.1–1.7 μm) [14, 15] (see Fig. S-2(b) in the ESM). Following injection of SWNTs, the vessels of the mouse were brightly fluorescent in the 1.1–1.4 μm range, as the SWNTs were circulating in the blood (Fig. S-2(c)). At 6-h post-injection (Fig. S-2(d)), tumor accumulation was already evident. After 1 day, the majority of SWNTs were cleared from the blood and the high tumor uptake became evident from the high tumor contrast in the NIR fluorescence images (Fig. 2(d), Figs. S-1(e) and S-1(f)), consistent with the *ex vivo* biodistribution data (Fig. 2(b)). Several dozen mice were imaged during the course of this study and all showed high tumor uptake of SWNTs which resulted in easy imaging of tumors by using the intrinsic NIR photoluminescence of nanotubes.

Additionally, we conducted confocal Raman imaging of thin (10 μm) slices of tumor tissues obtained 3 days post-injection of SWNTs. We used the characteristic graphitic (G) band (Fig. 2(f)) of the SWNT at ~1600 cm$^{-1}$ to map out the concentration of SWNTs at a given point in the tissue slice [11], allowing us to glean the exact distribution of SWNTs within the tumor with a spatial resolution of ~20 μm (Fig. 2(e)). The Raman imaging revealed an appreciable amount of SWNTs deep inside the tumor, in addition to SWNTs on the outer edge of the tumor (Fig. 2(e)). Raman imaging is another important feature of SWNTs owing to strong resonance Raman effects in one-dimensional systems [1].

## 2.3 Photothermal treatment of tumors with SWNTs and AuNRs

Twenty BALB/c mice were inoculated with one subcutaneous 4T1 tumor each over the right shoulder, directly beneath the skin. Five days after the inoculation, 10 of the mice were injected intravenously with 200 μL of a 2 μmol/L (0.35 mg/mL) solution of SWNTs with a 50% DSPE-mPEG/50% $C_{18}$-PMH-mPEG coating [5]. This was approximately a 3.6 mg/kg body weight dose of SWNTs. Three days after injection, all mice were irradiated for 5 min at 0.6 W/cm$^2$ with an 808 nm NIR laser with a laser spot size of 4.4 cm$^2$ (Fig. 3). The laser power was kept constant throughout the irradiation. This led to a rapid temperature rise in the first minute which then leveled to approximately 52–54 °C over the next four minutes. The temperature was monitored continuously by thermal imaging and checked periodically by a thermoprobe placed directly in contact with the tumor. The physical appearance of the tumor whitened. The tumors had an average size of ~22 mm$^3$ at the start of irradiation. Thermal images, taken at various time points during the heating, showed that tumors on non-injected mice heated up very slowly and reached an average temperature of 43 °C (Fig. 3(b)). In contrast, the tumors of SWNT-injected mice reached temperatures of 52.9 °C after 5 min of irradiation (Fig. 3(a)) under the same conditions ($n = 10$). Overall, while the laser spot size was much larger than the tumor (4.4 cm$^2$ as compared to a tumor size of less than 0.5 cm$^2$), thermal imaging indicated that the healthy, illuminated tissue of SWNT-injected mice remained closer to normal body temperatures (Fig. 3(a)). This confirms that SWNTs taken up by tumors were responsible for the absorption of laser light, and subsequent heating of the tumor tissue. After treatment, the tumors in the mice injected with SWNTs turned black. Eventually this led to an eschar forming over the top of the tumor.

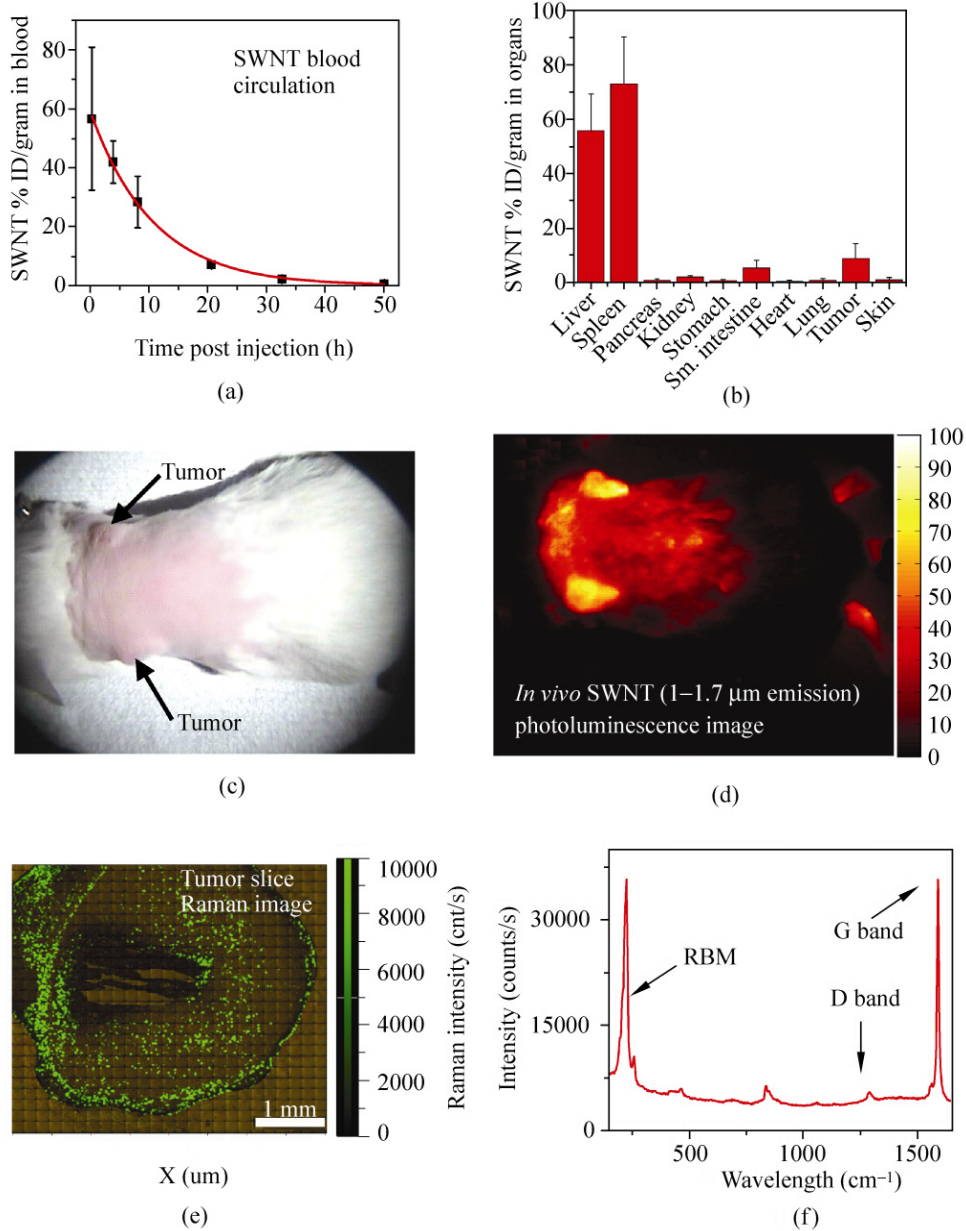

**Figure 2** *In vivo* NIR imaging using carbon nanotube photoluminescence. (a) SWNT blood circulation data. Black symbols correspond to experimental data. The red line is a first order exponential fit to the data, indicating a half-life of approximately 6.9 h. After 48 h, the SWNT signal had dropped below the detection limit. These data were obtained by Raman spectroscopy (see the Methods section). Error bars are based on three mice per group. (b) Biodistribution of SWNTs in various organs. Three mice were sacrificed 48 h after injection. The SWNT concentration was measured using Raman spectroscopy (see the Methods section). (c) Optical image of a BALB/c mouse with two 4T1 tumors (indicated by arrows). (d) An NIR PL image taken 48 hours post-injection. High tumor contrast is seen as the SWNTs are cleared from blood circulation, leaving SWNTs passively taken up in the tumors through the enhanced EPR effect. (e) A Raman image (the green color represents the intensity of the Raman G band of SWNTs) showing the distribution of SWNTs in a 10 μm thin slice of tumor. Mice were sacrificed three days post-injection and Raman mapping was performed using Raman spectroscopy with 20 μm step size (see the Methods section ). (f) A Raman spectrum of a SWNT solution. Note the strong characteristic G peak of at 1600 cm$^{-1}$ which is used for Raman imaging

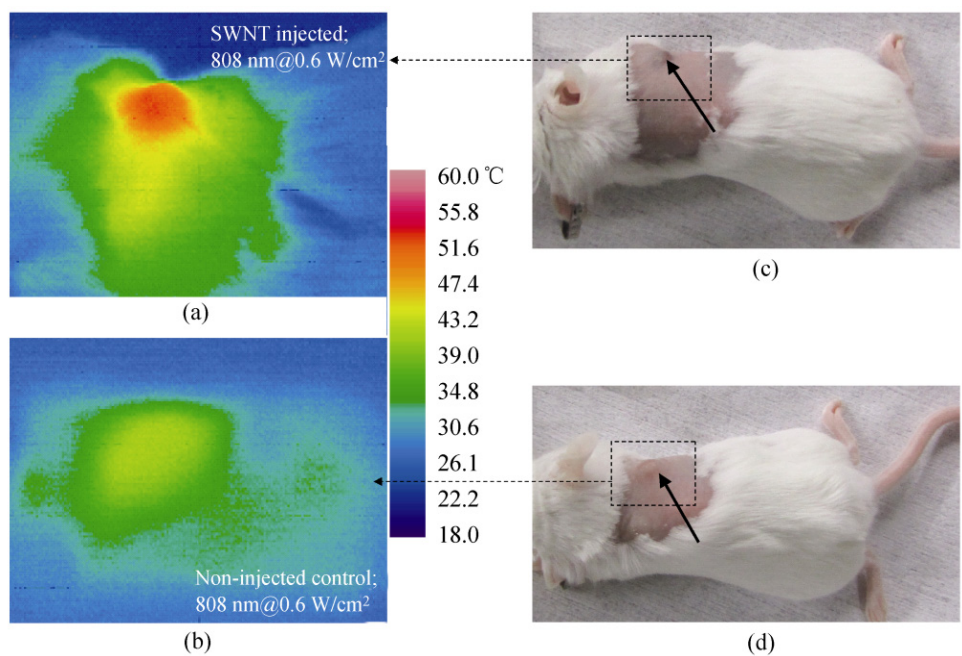

**Figure 3** Thermal imaging of a tumor during photothermal treatment. (a) A thermal image of a tumor-bearing mouse injected with 200 μL of 0.35 mg/mL (3.6 mg/kg) solubilized SWNT solution under 808 nm laser irradiation. The therrmal image was taken 4.5 min into NIR laser irradiation of the tumor and surrounding area at a power of 0.6 W/cm$^2$. Nine mice injected with solubilized SWNTs were thermally imaged while being irradiated, all showing similar results. The temperature rise in the skin surrounding the tumor was below the temperature threshold for tissue damage and no skin damage in the near or long term was observed. (b) A thermal image of a control tumor-bearing mouse without SWNT injection. The therrmal image was taken 4.5 min into 808 nm NIR laser irradiation of the tumor and surrounding area at a power of 0.6 W/cm$^2$. (c) Optical image of the BALB/c mouse in (a) injected with SWNTs, immediately before NIR laser irradiation. (d) Optical image of the mouse in (b) taken immediately before NIR laser irradiation. The arrows point to the subcutaneous tumor in the mouse

After the skin over the treated area on SWNT-injected, NIR irradiated mice had healed—which occurred on average about two weeks after the heating—no tumor remained. The only indication of any tumor was a small scar. In contrast, the mice in the control NIR radiation group without SWNT injection did not show any sign of tumor damage. All tumors in the control group continued to grow (Fig. 4(a)), and by the 36th day after heating, all mice in the control group were considered "non-survival" due to their large tumor size (>0.5 cm$^3$, Fig. 4(b)). In strong contrast, all ten of the mice in the SWNT-injected NIR laser treated group were considered surviving without any tumor re-growth ~6 months post-treatment (Figs. 4(a) and 4(b)).

For comparison, we examined the NIR photothermal heating capabilities of AuNRs with a longitudinal plasmon resonance near 808 nm (Figs. 1(e) and 1(f)). The absorption coefficient for SWNTs at 808 nm is 46.5 L/(g·cm) [38]. Based on the UV–Vis spectrum, the absorption coefficient for AuNRs at 808 nm is 13.89 L/(g·cm). Five mice bearing two tumors each were injected intravenously with 200 μL of the AuNRs at a dose of ~35 mg/kg, and the AuNRs were allowed to circulate and distribute for 48 h. We found that AuNRs, while able to heat and ablate tumors at the previously reported power levels of 2 W/cm$^2$ of 808 nm laser irradiation [27, 29] for 5 min, were not able to significantly heat or damage tumors when they were illuminated with NIR laser light with a power of only 0.6 W/cm$^2$ for 5 minutes (Figs. 5(a)–5(c)). The tumors heated at 2 W/cm$^2$ formed an eschar and diminished while the tumors heated at 0.6 W/cm$^2$ continued to grow unabated (Fig. 5(d)). Based on thermal imaging, the average temperature of tumors in mice that were injected with AuNRs and heated at 2 W/cm$^2$ for

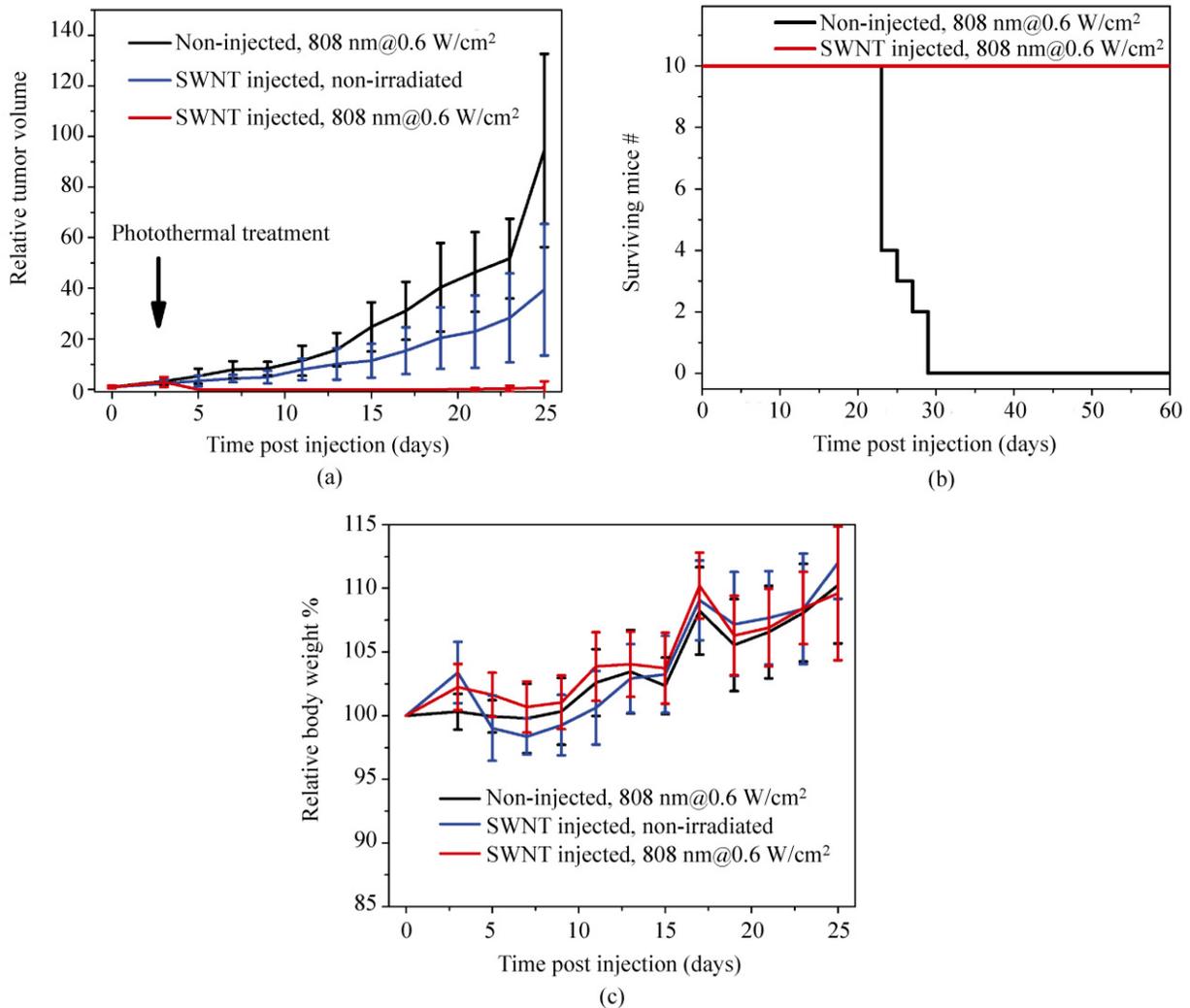

**Figure 4** Complete tumor elimination by photothermal treatment with SWNTs. (a) Plot of relative tumor volume vs. time for control group (10 mice) not injected with SWNTs but laser irradiated at 808 nm with 0.6 W/cm$^2$ power for 5 min, control group (four mice) that were injected with SWNTs but received no NIR laser irradiation, and treatment group (10 mice) injected with SWNTs and laser irradiated at 808 nm with 0.6 W/cm$^2$ power. Injection of 200 μL of 0.35 mg/mL (3.6 mg/kg) solubilized SWNT solution occurred on day 0 and NIR laser irradiation occurred on day 3. (b) Survival curve of control (no SWNT injection, laser irradiated at 808 nm with 0.6 W/cm$^2$ power) group (10 mice) vs. the treated (SWNT injection, laser irradiated at 808 nm with 0.6 W/cm$^2$ power) group (10 mice). Mice were deemed "non-survival" once the tumor volume exceeded 500 mm$^3$. No mice from the control group survived past 30 days post-treatment. All mice from the treatment group were surviving and tumor-free at the end of two months. (c) Normalized body weight for various mouse groups in (a). No mice in any group experienced significant weight loss (greater than 10%) during the study

5 min was much greater (~50 °C) than that of tumors heated at 0.6 W/cm$^2$ (~40 °C). Our data is consistent with the best result achieved previously by photo-ablation with AuNRs [26, 28], i.e., *in vivo* tumor destruction was also observed at a NIR laser power of ~2 W/cm$^2$ with a similar injected dose of AuNRs to that used here.

In our study, no mice that had been injected with SWNTs and exposed to NIR laser irradiation showed any weight loss or signs of distress (Fig. 4(c)). 48 days after photothermal treatment, blood was collected from the ten SWNT-treated mice and from five untreated, healthy control mice for analysis. All blood panel values for the treated mice fell within the previously reported normal range for healthy BALB/c mice [39] and matched well with those of healthy, untreated animals (see

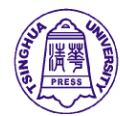

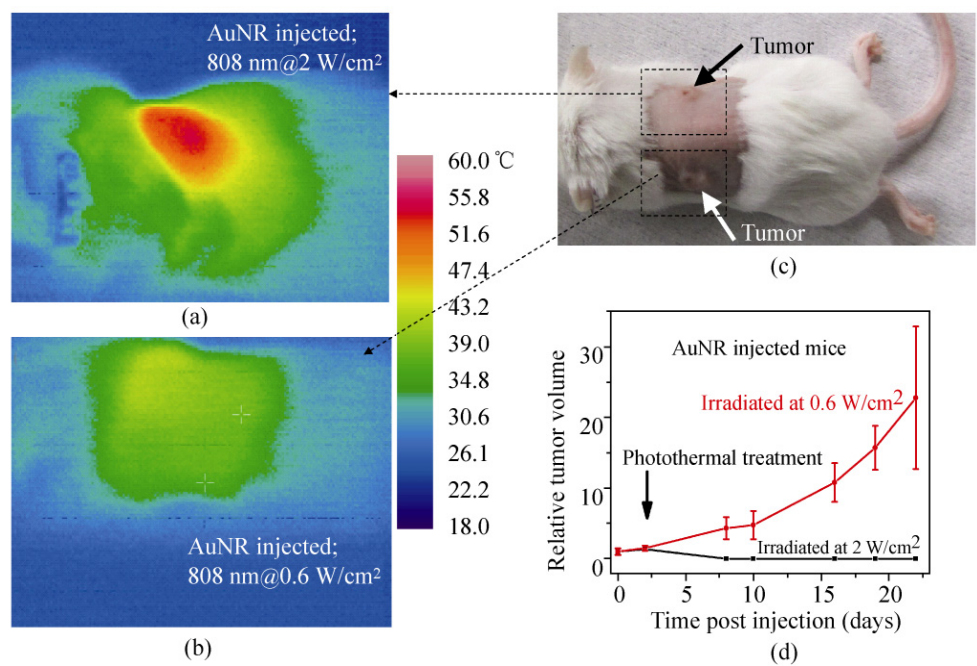

**Figure 5** In vivo photothermal treatment of tumors with gold nanorods: (a) a thermal image of a mouse bearing two tumors injected with 200 μL of 3.5 mg/mL (35 mg/kg) AuNR solution, taken of the left shoulder tumor 4.5 min into NIR laser irradiation at a power of 2 W/cm$^2$; (b) a thermal image of the same mouse as in (a) injected with 200 μL of 3.5 mg/mL (35 mg/kg) AuNR solution, taken of the right shoulder tumor 4.5 min into NIR laser irradiation at a power of 0.6 W/cm$^2$; (c) an optical image of the same BALB/c mouse as in (a) and (b) taken immediately before NIR laser irradiation; (d) plots of relative tumor volume vs. time for AuNR injected mice for tumors heated at 0.6 W/cm$^2$ for 5 min (five mice/group) and tumors heated at 2 W/cm$^2$ for 5 min (five mice/group)

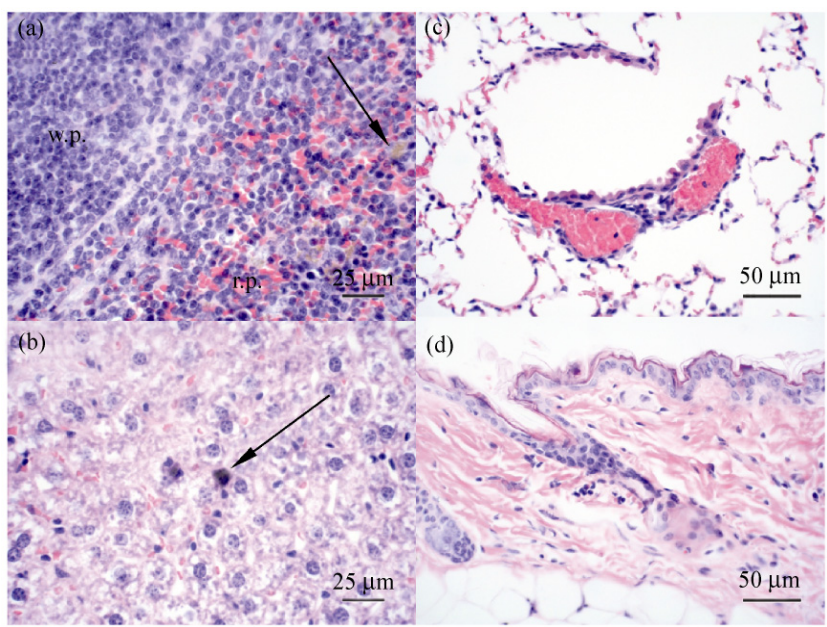

**Figure 6** Histology staining of the organs of mice injected with SWNTs after long-term exposure. Photomicrographs of hematoxylin and eosin (H&E)-stained (a) spleen, with white pulp (w. p.) and red pulp (r. p.); (b) liver; (c) lung; and (d) haired skin from mice 144 days after injection of SWNTs. Note that all organs appeared histologically normal. Notice the presence of dull to dark gray-brown, finely granular, intracytoplasmic pigment (indicated by black arrows) within scattered macrophages of the liver and spleen. This intracytoplasmic pigment is distinct from the variably-sized, brown to golden-brown hemosiderin pigment noted in other parts of the red pulp of the spleen, and is interpreted as residual SWNTs within macrophages with long life spans in the liver and spleen. The lack of organ or tissue damage observed by histologic evaluation, combined with the normal blood biochemical data and lack of obvious behavioral or body weight changes, suggests the overall non-toxic nature of the SWNTs

Table S-1 in the ESM). The excretion pathway for SWNTs solubilized by amphiphilic polymers has previously been shown to be through the biliary pathway [38]. Our biodistribution data showed high signals of SWNTs in the liver as well as in the intestine 48 h after injection (Fig. 2(b)), consistent with fecal excretion of SWNTs [38]. In order to ascertain the long term toxicity of the solubilized SWNTs, three mice from the treatment group were submitted for histology 144 days after injection of SWNTs (Fig. 6) Additionally, a blood chemistry panel was performed on five mice 163 days post-injection (see Table S-1 in the ESM). Both histology and blood chemistry data suggested healthy normal functions of the treated mice.

## 3. Discussion

Our SWNTs (average length~140 nm), with appropriate chemical functionalization, afforded high passive accumulation in tumors relative to surrounding organs and provide an opportunity to combine imaging and photothermal therapy of tumors by utilizing the intrinsic optical properties of nanotubes. The inherent characteristic Raman bands and NIR PL of SWNTs allow for both *in vivo* and *ex vivo* tracking and imaging. Monitoring the fluorescence of the SWNTs gives unambiguous confirmation of SWNT uptake by the tumors. The fact that photoluminescence of SWNTs in the tumor was clearly present and can be used for tumor imaging suggests a lack of significant aggregation of nanotubes in the tumor, since this would cause quenching of the photoluminescence of the nanotubes [36]. Whole body NIR fluorescence imaging allows visualization of the preferential tumor accumulation of SWNTs (Fig. 2(d), Fig. S-2 in the ESM). SWNTs are interesting as NIR fluorescent probes due to their unusual emission range of 1.0–1.4 μm (Fig. 1(b)), which is ideal for biological imaging due to the inherently low autofluorescence in this region [14, 15]. The large Stokes shift between excitation and emission bands of SWNTs allows for excitation in the biological transparency window near 800 nm, while detecting and imaging with reduced background from autofluorescence and scattering in the 1.0–1.4 μm region [14, 15]. Also, NIR fluorophores with emission in the 1.0–1.4 μm range have higher tissue penetration than those with emission near 800 nm, considering the effects of scattering by tissue. Thus, despite relatively low PL quantum yield of SWNTs (<1%–2%) [14, 15], the ultralow imaging background and sufficient tissue penetration afford excellent contrast for SWNT photoluminescence agents when imaging subcutaneous tumors in mice.

The depth penetration of the NIR laser does become an issue when treating tumors deep within the body. Since the penetration depth is less than 1 cm, other methods besides an external laser source must be devised in order to effectively treat deep body tumors. This could be resolved through the use of an internal fiber optic laser source threaded into the body [40].

When the NIR fluorescence images show a high signal in the tumor and a low signal in the surrounding tissue, we are able to heat the tumors without incurring damage to healthy tissue. Never before have SWNTs been intravenously (i. v.) injected, imaged in the tumor by their intrinsic photoluminescence and subsequently heated by NIR irradiation. To demonstrate the effectiveness and benefit of SWNTs relative to other materials, we compared the photothermal therapy capability of commonly used AuNRs with that of SWNTs. At the same mass concentration of 0.35 mg/mL, SWNTs exhibited an optical absorbance at 808 nm which was 328% greater than that of AuNRs (Fig. 1(f)). The AuNRs exhibited a strong optical absorption at 808 nm due to plasmonic resonance; the wavelength of this absorption is tunable by varying the aspect ratio and size [41]. The high NIR absorbance of SWNTs originates from their one-dimensional electronic structures. Note that the SWNTs used here contain many different chiralities, each of which has peaks of absorption and emission at different wavelengths [36] (Fig. 1(b)). Therefore, only a few chiralities of SWNTs in our current sample have an absorption peak at 808 nm in resonance with the 808 nm laser. With advances in chirality separation techniques, we will be able to use a selected single-chirality SWNT with greatly increased optical absorbance to afford higher photothermal efficiency at a further reduced injection dose. With chirality-separated materials, tumor targeting to increase tumor uptake, and optimal surface coating for long blood circulation and high tumor uptake, the injection dose needed for SWNTs to achieve tumor ablation could be further lowered from the currently lowest reported level of 3.6 mg/kg demonstrated here.

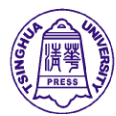

It should also be noted that compared to previous literature results, our SWNT photothermal treatment affords highly effective tumor destruction at much lower injected doses and laser irradiation power than other photothermal therapy agents reported, including gold nanoshells, nanorods and multi-walled nanotubes (see Table S-2 in the ESM). For SWNTs, 100% tumor destruction was achieved with a dose which was only 10%, and a laser irradiation power which was only 30%, of the corresponding values required to achieve 100% tumor destruction with Au nanorods It should be noted that the laser irradiation power (0.6 W/cm$^2$) and exposure time used here for SWNTs are slightly above the laser safety standards for humans. The maximum skin exposure for a continuous wave 808 nm laser, based on the American National Standard for the Safe Use of Lasers, is ~0.33 W/cm$^2$.

Both AuNRs [26] and SWNTs when properly functionalized have long blood circulation times and upon intravenous injection accumulate in tumors through the EPR effect. For both materials, highly hydrophilic poly(ethylene) glycol (PEG)-based polymers are used as a "masking" device *in vivo*, coating both AuNRs and SWNTs to avoid rapid attack by the body's immune system. Toxicity and long term retention/excretion is a concern for all nanomaterials *in vivo*. The extent of acute or long-term toxicity of carbon nanotubes is currently under investigation. Our short SWNTs coated by phospholipids have high solubility and stability against agglomeration in aqueous solutions and serum and have been demonstrated to be biocompatible and non-toxic to mice over a monitoring period of up to six months [38, 42].

Stable PEGylated SWNTs do not seem to damage mice organs after i. v. injection and the majority appear to be excreted from the body over the course of < two months [38, 42]. Results from the current work (Table S-1 in the ESM) further confirm non-toxicity of our SWNTs *in vivo*. Histology was conducted on the vital organs of the mice in the treatment group 144 days after injection of solubilized SWNTs (Fig. 6). No pathology or cancer was observed and the organs were all in healthy condition. Small amount of SWNTs appeared to be still retained in the macrophage cells of the liver and spleen as shown by the dull grey pigmentation seen in stained slices of these organs (Figs. 6(b) and 6(c)). However, no damage was seen in any of the tissues examined and the residual SWNTs did not cause any loss of function of the spleen or liver. Furthermore after observation for 144 days, despite the highly metastatic nature of 4T1 tumors, no metastases were observed in any of the mice in the treatment group. Based on the biodistribution data (Fig. 2(b)), as long as NIR light irradiation avoids the *reticuloendothelial* system (RES) organs (including liver and spleen that have high SWNT uptake) during tumor irradiation, few side effects and little organ damage is expected.

Longer term effects should be investigated (currently underway in our laboratory) to ensure the complete biocompatibility of PEGylated SWNTs. Note that possible toxicity effects of gold nanomaterials have also been under examination. Gold nanoshell photothermal cancer treatment is already in Phase I clinical trials [43]. AuNRs coated by CTAB, while toxic *in vivo* and *in vitro*, are non-toxic when exchanged with more biocompatible coatings [44]. AuNRs exchanged from CTAB to a mPEG-SH coating have been shown to be non-toxic in the short term *in vivo* [26, 28]. To our knowledge, the long term fate, excretion, and toxicity of intravenously injected gold nanomaterials have not yet been reported. The long term fate and toxic effects for various nanomaterials require systematic investigation. The benefits and risks of these materials for potential future clinic use can then be assessed.

## 4. Conclusions

*In vivo* photothermal laser heating of tumors coupled with the ability to image and visualize their uptake into tumors makes SWNTs a promising potential material for future photothermal treatment. The novelty of the current work lies in several areas. On the imaging front, we have been able to achieve high tumor uptake of SWNTs, and used the intrinsic NIR PL of SWNTs for tumor imaging for the first time. With regards to treatment, we took advantage of the high optical absorbance of SWNTs in the NIR to heat tumor masses to the point of complete cell death, using lower power (0.6 W/cm$^2$) than previously reported at injection doses as low as 3.6 mg/kg. The ability to faithfully track the tumor distribution of SWNTs from their Raman signal is also an important feature of SWNTs. In order to put the efficiency of SWNTs in perspective, we conducted a comparative study with AuNRs. Further improvement, such as chiral separation of SWNTs, could lead to a significantly lower injection dose and lower laser powers needed for tumor destruction. Continued investigation of the toxicity, if any, of well-coated SWNTs employed in cancer therapy is important as well. Nevertheless, the ability to image SWNTs within the body of live mice and the high performance of SWNTs in photothermal therapy for tumor elimination lays the groundwork for future experimentation into using the inherent properties of SWNTs for cancer detection and treatment.

# 5. Methods

## 5.1 Materials

HiPco single-walled carbon nanotubes were obtained from Carbon Nanotechnologies Inc. Poly(maleic anhydride-*alt*-1-octadecene) (molecular weight 30 to 50 kDa) was purchased from Sigma-Aldrich. Both mPEG-NH$_2$ and DSPE-mPEG were obtained from Laysan Bio Inc. Regenerated cellulose dialysis membrane bags were obtained from Fischer Scientific.

## 5.2 Synthesis of C$_{18}$-PMH-mPEG

Polymer C$_{18}$-PMH-mPEG was synthesized in the following manner based on previous work [34]. Methoxy-poly(ethylene glycol)-amine (285.7 mg, 0.05714 mmol, mPEG-NH$_2$, 5 kDa) was combined with poly(maleic anhydride-*alt*-1-octadecene) (10 mg, 0.0286 mmol) in 15 mL of a 9:1 DMSO/pyridine mixture. The solution was allowed to stir for 12 h at room temperature, followed by the addition of 1-ethyl-3-(3-dimethylaminopropyl) carbodiimide hydrochloride (21.8 mg, 0.11 mmol) (EDC·HCl). The reaction was continued for 24 h, followed by dialysis to remove excess mPEG-NH$_2$.

## 5.3 Preparation of SWNT suspensions

A 50% DSPE-mPEG/50% C$_{18}$-PMH-mPEG SWNT nanotube solution was prepared by combining 0.2 mg/mL of HiPco tubes with 0.6 mg/mL of DSPE-mPEG and 0.6 mg/mL of C$_{18}$-PMH-mPEG in 30 mL of water. The solution was sonicated for 1 h followed by centrifugation (6 h, 22 000 $g$) to remove any bundles or aggregates. The resulting supernatant was collected and filtered eight times through a 100 kDa pore size filter (Millipore) to remove excess polymer. 200 μL solutions of 2 μmol/L SWNT were prepared in 2× *phosphate-buffered saline* (PBS). This was done by adjusting the concentration based on the absorption peak at 808 nm having an extinction coefficient [5] of 7.9 × 10$^6$ L/mol cm.

## 5.4 Synthesis of gold nanorods (AuNRs)

Gold nanorods (AuNRs) were prepared via seed-mediated growth in cetyltrimethylammonium bromide (CTAB, Sigma-Aldrich) as described previously [37] and reacted with methyoxy-terminated, thiolated poly(ethylene glycol) (mPEG-SH, 5 kDa, Laysan Bio) to provide biocompatible AuNRs with a longitudinal surface plasmon centered at ~800 nm. 5 mL of 0.2 mol/L CTAB was mixed with 5 mL of 0.5 mmol/L HAuCl$_4$ (Sigma-Aldrich, 99.9%), and the Au(III) was reduced to form seed particles with 600 μL of ice-cold 10 mmol/L NaBH$_4$, yielding a yellow–brown solution. The AuNR growth solution was prepared by adding 80 mL of 0.2 mol/L CTAB to 3.2 mL of 4 mmol/L AgNO$_3$, followed by addition of 80 mL of 1 mmol/L HAuCl$_4$ and finally 1.12 mL of 78.8 mmol/L ascorbic acid, yielding a colorless solution. 192 μL of seed solution was added to the growth solution, facilitating AuNR growth over the course of several hours at 25 °C. To exchange the CTAB coating by mPEG-SH, 40 mL of as-made AuNRs were centrifuged at 22 000 $g$ for 7 min and the supernatant was discarded. The AuNRs were resuspended in 15 mL of deionized water, and mPEG-SH was added to give a final concentration of 250 μmol/L. After gentle agitation at room temperature for one hour, the solution was added to a pre-soaked regenerated cellulose dialysis membrane (MWCO 3500 Da, Fisher), and dialyzed against 5 × 1 L deionized water. Excess mPEG-SH and any residual

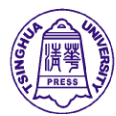

CTAB were then removed by 4× centrifugal filtration in water (Millipore Amicon Ultra, 30 kDa MWCO), which also allowed concentration of the AuNR sample prior to injection.

### 5.5 Photoluminescence excitation and emission Spectra (PLE)

PLE spectra were recorded on a homebuilt NIR spectroscopy setup. The excitation source was a 150 W ozone-free xenon lamp (Oriel) which was dispersed by a monochromator (Oriel) to produce excitation lines with a 15 nm bandwidth. The excitation light was focused onto a 1 mm quartz cuvette containing the sample. Emission was collected in transmission geometry. The excitation light was rejected using an 850 nm long-pass filter (Omega). The emitted light was directed into a spectrometer (Acton SP2300i) equipped with a liquid nitrogen cooled InGaAs linear array detector (Princeton OMA-V). Spectra were corrected post-collection to account for the sensitivity of the detector and the power of the excitation.

### 5.6 Mouse handling and injection for imaging

Female BALB/c mice obtained from Charles Rivers were housed at Stanford Research Animal Facility (RAF) under Stanford Institutional Animal Care and Use Committee (IACUC) protocols. Six mice had their backs shaved and were injected bilaterally with ~2 million 4T1 tumor cells on both shoulders. The tumors were allowed to grow for 5 days to a size approxi- mately 3–6 mm in length and width. After this time, the mice were anesthetized with isoflurane gas and an optical image along with a NIR PL image was taken. A solubilized SWNT solution was tail vein (intravenously) injected into the six mice and NIR PL images were collected over the following 48 h. Three of the mice were used for circulation and biodistribution tests.

### 5.7 Mouse handling and injection for treatment

Female BALB/c mice obtained from Charles Rivers were housed at Stanford Research Animal Facility under Stanford Institutional Animal Care and Use Committee protocols. Twenty mice had their backs shaved and were injected bilaterally with ~2 million 4T1 tumor cells on their right shoulders. Four mice had their backs shaved and were injected bilaterally with ~2 million 4T1 tumor cells on both of their shoulders.

### 5.8 NIR PL Imaging

NIR PL images were collected using a 2-D liquid nitrogen cooled InGaAs detector (Princeton). A 20 W 808 nm diode laser (RPMC) was used as the excitation source. The excitation power density at the imaging plane was 0.15 W/cm$^2$. The excitation was shuttered so that the mice were only subjected to the excitation beam for the duration of the exposure time. Exposure time for all images shown was 300 ms.

### 5.9 Blood Circulation

Three mice were intravenously injected with 200 μL solutions of 2 μmol/L (0.35 mg/mL) SWNTs. At given time points, approximately 4 μL of blood was drawn from the tail of each mouse. The blood was mixed with 8 μL of tissue lysis buffer solution (1% SDS, 1% Triton X-100, 40 mmol/L Tris acetate, 10 mmol/L EDTA, 10 mmol/L DTT). The height of the characteristic G peak in the Raman spectrum of the SWNTs corresponds directly to the concentration of SWNTs. This, when compared with the injected solution, allows us to quantify the amount of SWNTs still circulating in the blood as previously reported [38].

### 5.10 Biodistribution

The mice used for circulation tests were sacrificed at 48 h post-injection. Organs were collected and weighed, then dissolved in a known amount of tissue lysis buffer. The mixtures were made homogenous through heating and blending. After this, Raman spectroscopy was used in the same manner as for the blood circulation measurements to identify the concentration of SWNTs in the various organelles, as has been reported previously [38].

### 5.11 SWNT distribution in the tumors

Three mice that had not been used in any other tests were inoculated with two 4T1 tumors bilaterally. Five days after inoculation, 200 μL of a solution of 2 μmol/L (0.35 mg/mL) of solubilized SWNTs was injected. 72 h after injection, the mice were sacrificed and tumors were collected and frozen in Tissue-Tek OCT compound. The tumors were sliced into 10 μm thin layers and placed on quartz. Raman (LabRam 800 Horiba Jobin Yvon) mapping using 785 nm excitation was conducted with 20 μm steps, 0.1 s integration time, and five accumulations per point with an average laser spot size of ~1 μm$^2$.

### 5.12  NIR laser irradiation of tumors in mice

Twenty four mice in total were included in the mouse heating study. Twenty mice had one tumor each while four mice had two tumors each. Ten of the twenty mice with one tumor were intravenously injected with 200 μL aliquots of 2 μmol/L solubilized SWNTs five days after tumor inoculation. The four mice with two tumors each were intravenously injected with 200 μL of 2 μmol/L of solubilized SWNTs five days after tumor inoculation. Three days post-injection, all twenty four mice were anesthetized with isoflurane gas and exposed to 808 nm collimated laser light. The light was shone directly on the right shoulder for 5 min on a 4.4 cm$^2$ sized spot with a power of 0.6 W/cm$^2$ (the four mice with bilateral tumors did not have any irradiation on the tumor located on the left shoulder). After this time, mice were monitored by body weight and tumor size every two days for the following 35 days (or until the tumor size exceeded the survival size of 0.5 cm$^3$). Five mice with two tumors each (5 days post inoculation) were injected with 200 μL of 3.5 mg/mL AuNR solution (35 mg/kg injection dose). 48 h after injection, the tumor on the right shoulder was irradiated with 808 nm laser light at a power of 0.6 W/cm$^2$ for 5 min. Afterwards, the tumor on the left shoulder was irradiated with 808 nm laser light at a power of 2 W/cm$^2$ for 5 min. Tumor growth was monitored for several weeks or until the tumor reached 0.5 cm$^3$ in volume.

### 5.13  Thermal imaging

Thermal images were taken using a MikroShot camera (Mikron). During heating, thermal images were collected for mice before heating, and at 1, 2, and 5 minutes into the heating process. Throughout heating, the temperature of the mouse tumors was monitored by the MikroShot camera.

### 5.14  Histology and blood chemistry

51 days after injection of SWNTs, blood was collected from ten injected/irradiated mice and five control mice. Blood was again collected from five injected/irradiated mice 163 days after injection of SWNTs. Blood was collected by retro-orbital bleeding and serum chemistry analyzed by the Diagnostic Laboratory, Veterinary Service Center, Department of Comparative Medicine, Stanford University School of Medicine. 144 days after injection of SWNTs, three mice that had been injected/irradiated were submitted for histology. A full necropsy was performed, and all internal organs were harvested, fixed in 10% neutral buffered formalin, processed routinely into paraffin, sectioned at 4 mm and stained with hematoxylin and eosin (H&E). The following tissues were examined by optical microscopy: liver, kidneys, spleen, heart, salivary gland, lung, trachea, esophagus, thymus, reproductive tract, urinary bladder, eyes, lymph nodes, brain, thyroid gland, adrenal gland, gastrointestinal tract, pancreas, bone marrow, skeletal muscle, nasal cavities, middle ear, vertebrae, spinal cord, and peripheral nerves.

## Acknowledgements

This work was supported by Ensysce Biosciences, CCNE-TR at Stanford University and NIH-NCI RO1 CA135109-02.

**Electronic Supplementary Material**: Supplementary material detailing other properties of the photothermal agents, the complete time course of whole animal imaging, and blood chemistry data is available in the online version of this article at http://dx.doi.org/10.1007/s12274-010-0045-1 and is accessible free of charge.

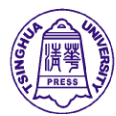

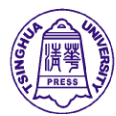